\documentclass[3p,number]{elsarticle}
\makeatletter
\def\ps@pprintTitle{%
	 \let\@oddhead\@empty
	 \let\@evenhead\@empty
     \def\@oddfoot{\footnotesize\itshape
	   Preprint \ifx\@journal\@empty
	   \else\@journal\fi\hfill\today}%
	 \let\@evenfoot\@oddfoot}
\makeatother
\usepackage{hyperref}
\usepackage{etex}
\usepackage{amsmath}
\usepackage{amssymb}
\usepackage{amsthm}
\usepackage{array}
\usepackage{booktabs}
\usepackage{graphicx}
\usepackage{calc}
\usepackage{url}
\usepackage{multirow}
\usepackage{wrapfig}
\usepackage{tikz}
\usetikzlibrary{arrows,automata,shapes,shapes.multipart,decorations.markings,positioning}
\usepackage[all]{xy}
\usepackage[trim]{tokenizer}
\usepackage[labelformat=simple]{subcaption}

\newcommand{\net}{\mathsf{net}}
\newcommand{\word}{\mathsf{word}}
\newcommand{\sent}[1]{\ensuremath{\mathtt{#1}}}

\newcommand\tuple[1]{\langle #1 \rangle}
\newcommand{\sset}[2]{\left\{~#1  \left|
      \begin{array}{l}#2\end{array}
    \right.     \right\}}

\renewcommand{\vec}{\bar}

\newcommand{\ND}{\mathit{noneDown}}
\newcommand{\NU}{\mathit{noneUp}}
\newcommand{\OD}{\mathit{oneDown}}
\newcommand{\OU}{\mathit{oneUp}}
\newcommand{\once}{\mathit{once}}
\newcommand{\sorts}{\mathit{sorts}}
\newcommand{\update}{\mathit{update}}
\newcommand{\used}{\mathit{used}}
\newcommand{\valid}{\mathit{valid}}
\newcommand{\less}{\mathit{less}}

\newcounter{sncolumncounter}
\newcounter{snrowcounter}

\newtheorem{lemma}{Lemma}
\newtheorem{definition}{Definition}
\newtheorem{theorem}{Theorem}

\newcommand{\sncolwidth}{0.7} %

\newcommand{\addarrow}[5][0.75]{
  \draw[>=stealth,postaction={decorate},line width=1.5 pt,decoration={markings,mark=at position #1 with {\arrow[scale=1.2]{>}}}] (\sncolwidth*#2,#3)--(\sncolwidth*#4,#5);
}
\def \nodeconnectionref#1{%
  \foreach \i in {#1}{%
    \GetTokens{nodesrc}{nodedest}{\i}
    \draw (\value{sncolumncounter}*0.7,\nodesrc+1) node[circle,inner sep=0pt,minimum size=3pt,fill=black]{}--(\value{sncolumncounter}*0.7,\nodedest+1) node[circle,inner sep=0pt,minimum size=3pt,fill=black]{};
  }
  \addtocounter{sncolumncounter}{1}
}
\def \nodeconnection#1{%
	\foreach \i in {#1}{%
		\GetTokens{nodesrc}{nodedest}{\i}
		\draw (\value{sncolumncounter}*0.7,\value{snrowcounter}-\nodesrc) node[circle,inner sep=0pt,minimum size=3pt,fill=black]{}--(\value{sncolumncounter}*0.7,\value{snrowcounter}-\nodedest) node[circle,inner sep=0pt,minimum size=3pt,fill=black]{};
	}
	\addtocounter{sncolumncounter}{1}
}
\newenvironment{sortingnetwork}[2]
{
  \setcounter{sncolumncounter}{0}
  \setcounter{snrowcounter}{#1}
  \def \sn@fullsize{15}
  \begin{tikzpicture}[scale=#2*0.7]
}
{
  \foreach \i in {1, ..., \value{snrowcounter}}
  {
    \draw (-\sncolwidth,\i)--(\sncolwidth*\value{sncolumncounter}+\sncolwidth,\i);
  }
  \end{tikzpicture}
}
\makeatother

\begin{document}

\title{Joint Size and Depth Optimization of Sorting Networks\tnoteref{t1}} 

\tnotetext[t1]{Supported by the Spanish MINECO project TEC2015-69266-P (FEDER, UE)}  

\author{Jos\'e A. R. Fonollosa} \ead{jose.fonollosa@upc.edu} 

\address{Department of Signal Theory and Communications, Universitat Polit\`ecnica de Catalunya, Barcelona, Spain}

\begin{abstract}
Sorting networks are oblivious sorting algorithms with many interesting theoretical properties and practical applications. One of the related classical challenges is the search of optimal networks respect to size (number of comparators) of depth (number of layers). However, up to our knowledge, the joint size-depth optimality of small sorting networks has not been addressed before. This paper presents size-depth optimality results for networks up to 12 channels. Our results show that there are sorting networks for $n\leq9$ inputs that are optimal in both size and depth, but this is not the case for $10$ and $12$ channels. For $n=10$ inputs, we were able to proof that optimal-depth optimal sorting networks with 7 layers require 31 comparators while optimal-size networks with 29 comparators need 8 layers. For $n=11$ inputs we show that networks with 8 or 9 layers require at least 35 comparators (the best known upper bound for the minimal size). And for networks with $n=12$ inputs and 8 layers we need 40 comparators, while for 9 layers the best known size is 39.
\end{abstract}

\maketitle

\section{Introduction}
\label{sec:intro}
A sorting algorithm is data-independent or oblivious if the sequence of comparisons does not depend on the input list. Sorting networks are oblivious sorting algorithms with many practical applications and rich theoretical properties \cite{knuth1998art}. From the practical point of view, sorting networks are the usual choice for simple parallel implementations in both hardware and software such as Graphics Processing Units (GPUs). Moreover, sorting networks are also of interest for secure computing methods like secure multi-party computation, circuit garbling and homomorphic encryption \cite{bogdanov2014practical}. Other applications include median filtering, switching circuits, and encoding cardinality constraints in propositional satisfiability problems (SAT)\cite{abio2013parametric}. Interestingly, we use this cardinality constraint in this paper to perform the joint size-depth optimization of sorting networks in a SAT framework.

\begin{figure}[htb]
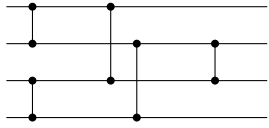

	\centering
	\begin{sortingnetwork}{4}{0.7}
		\nodeconnection{ {0, 1}, {2, 3}}
		\addtocounter{sncolumncounter}{2}
		\nodeconnection{ {0, 2}}
		\nodeconnection{ {1, 3}}
		\addtocounter{sncolumncounter}{2}
		\nodeconnection{ {1, 2}}
	\end{sortingnetwork}
	\caption{Comparator network of depth $3$ with $5$ comparators.}
	\label{fig:example}
\end{figure}

From the theoretical point of view comparator networks can be studied using the combinatorial and algebraic properties of permutations \cite{bruijn1974,bruijn1983}, as well as constrained boolean monotone circuits using the zero-one principle \cite[p.~223]{knuth1998art}. In the usual representation the $n$ input values are fed into networks of $n$ channels connected by comparators that swap unordered inputs from two channels. The sequence of data-independent comparisons can be parallelized grouping independent comparators in layers. The depth of a comparator network is the number of layers, i.e., the delay in a parallel implementation. The typical graphical representation of a comparator network is depicted in Figure \ref{fig:example}.

In this work we center our attention in the search of optimal sorting networks in both size (number of comparators) and depth (number of layers) for small values of $n$. For large values of $n$, Ajtai, Koml\'os, and Szermer\'edi \cite{ajtai1983sorting} presented in 1983 a method to construct sorting networks that have asymptotically optimal size and depth with $O(n \log n)$ comparators in $O(\log n)$ layers. However, despite this good asymptotic behavior, the original AKS network and other more recent variants, are currently of little practical interest because of the huge constant hidden in the big-$O$ notation. Simple recursive generation algorithms as the proposed by Batcher \cite{batcher1968sorting} provide much better sorting networks in term of size and depth for networks of practical interest. 

For small, fixed values of input $n$, the search for efficient sorting networks in terms of size or depth is a optimization quest with more than 50 years of history. For $n\leq16$ the best known networks in terms of size are more than 40 years old \cite{knuth1998art} but their optimality has only been recently proved for $n\leq10$ \cite{codish2016sorting}. The optimal-depth search follows a similar pattern. The best sorting networks in terms of depth for $n\leq16$ were already known 40 years ago \cite{knuth1998art} but the optimality proofs had to wait until 1989 $n\leq10$ \cite{parberry1989computer},  2013 for $n\leq16$ \cite{bundala2014optimal-depth} and 2015 for $n=17$ \cite{codish2016sorting}.

Recent optimality results are based on reducing the exponential complexity of a exhaustive search exploiting symmetries and efficient SAT encodings. In this paper we adapt optimal-depth SAT encodings \cite{bundala2014optimal-depth, ehlers2015new, codish2016sorting} to include additional cardinality constraints that limit the total number of comparators. We also need to reconsider some of the standard simplifications or restrictions for optimal-depth search problems that are not longer valid for joint size and depth optimizations. For example, we can no longer assume that the first layer of the network is of maximal-size.

A powerful symmetry-breaking tool in sorting network optimization is the use of a reduced set of prefixes that fix the first layers of the network reducing both the search space and the number of inputs of the sorting test. Bundala et al. developed in \cite{bundala2014optimal-depth} an efficient generation of complete sets of two-layer prefixes on $n$ channels for the specific optimal-depth problem. In our search for depth-size optimality results, we extend that work to consider any kind of two layer networks. Section \ref{sec:symbolic} covers the symbolic generation of complete sets of two layers prefixes modulo symmetry for this general case.

A recent related work of practical interest uses an evolutionary approach \cite{sorterhunter} to search for small and low depth sorting networks, but the search strategy lacks completeness. Our SAT-based approach is able to provide similar networks with provably size-depth optimality for small networks.

\section{Preliminaries}
\label{sec:Preliminaries}
A comparator network is a set of channels connected by a sequence of comparators as illustrated in Figure \ref{fig:example}. Channels are depicted as horizontal lines (with the first channel at the top). Each comparator (i,j) compares the input values ($in_i$, $in_j$) of the two connected channels $(1 \le i < j \le n)$ and if necessary rearrange them such that $out_i=min(in_i, in_j)$ and $out_j=max(in_i, max_j)$. The sequence of comparators can be grouped in maximal sets of independent comparators (layers) whose output can be computed in parallel. The depth of a comparator network is the number of layers. A sorting network is a comparator network that sorts all input sequences.

A key tool for the proof of correctness of sorting networks is the 0-1-principle \cite{knuth1998art}: if a sorting network for n channels sorts all $2^n$ sequence of 0's and 1's, then it sorts every arbitrary sequence of values.

Another important well-known fact is that any permutation of the input channels of a comparator network (followed by an untangling procedure) does not change its sorting properties \cite{knuth1998art}. Parberry \cite{parberry1989computer} use this property to fix the first layer in the search of optimal-depth lower bounds. He also mentions that we do not need to consider equivalent second layers up to permutation of channels. More recently, Bundala et al. \cite{bundala2014optimal-depth} studied the characterization and symbolic representation of equivalent two layer prefixes up to permutation for the same optimal-depth search problem. Two networks $C$ and $C'$ are equivalent up to permutation, denoted by $C \approx C'$, if there is a permutation $\pi$ such that $C'$ equals $\pi(C)$ (after the untangling procedure).

\section{Propositional Encodings for Joint Size and Depth Optimization of Sorting Networks}
\label{sec:SAT}

In this section we adapt the SAT encoding proposed by Codish et al. \cite{codish2016sorting} including additional cardinality constraints that limit the total number of comparators. For completeness' sake we include the original encoding and we follow a similar notation.

A comparator network of depth $d$ on $n$ channels is represented by a set of Boolean variables $C^d_n=\sset{g^k_{i,j}}{1 \leq i < j \leq n,1 \leq k \leq d}$, the value of $g^k_{i,j}$ indicating whether there is a comparator on channels $i$ and $j$ in layer $k$ in the network or not. 

\subsection{Validity encodings}
In a valid network the comparators of each layer are independent, i.e., each channel may be used only once:

\begin{align*}
\once^k_i(C^d_n) =& \bigwedge_{1 \leq i \neq j \neq l \leq n} \left( \neg g^k_{\min(i, j), \max(i, j) } \vee \neg g^k_{\min(i, l), \max(i, l) } \right) \\
\valid(C^d_n) =& \bigwedge_{1 \leq k \leq d, 1 \leq 1 \leq n} \once^k_i(C^d_n)
\end{align*}

\subsection{Sorting encodings}
We denote the inputs into the network by the variables~$v^0_i$, with $1 \leq i \leq n$, the variables $v^k_i$, with $1
\leq i \leq n$ and $1 \leq k \leq d$, store the value on channel $i$ in the network after layer $k$. A valid networks sorts if the following SAT constraints on $v^k_i$ $g^k_{i,j}$ are satisfied:

\begin{align*}
\used^k_i(C^d_n) =& \bigvee_{j <i} g^k_{j, i} \vee \bigvee_{i < j} g^k_{i, j} \\
\update^k_i(C^d_n) =& \left( \neg \used_i^k(C^d_n) \rightarrow (v^k_i \leftrightarrow v^{k-1}_i) \right) \wedge \\
&\bigwedge_{1 \leq j < i} \left( g^k_{j,i} \rightarrow \left(v^k_i \leftrightarrow (v^{k-1}_j \vee v^{k-1}_i )\right) \right) \wedge \\
&\bigwedge_{i < j \leq n} \left( g^k_{i,j} \rightarrow \left(v^k_i \leftrightarrow (v^{k-1}_j \wedge v^{k-1}_i )\right) \right)
\end{align*}

The $update$ constraint describes the impact of comparators on the values $v^k_i$ stored on each channel after every layer and the following $sorts$ equation includes all the $update$ constraints to assure the output $y$ for a specific input $x$, where in our case $y$ is the sorted version $x$.

\begin{align*}
\sorts(C^d_n, x) = \bigwedge_{1 \leq i \leq n} (v^0_i \leftrightarrow x_i) \wedge \bigwedge_{1 \leq k \leq d, 1 \leq i \leq n} update^k_i(C^d_n) \wedge \bigwedge_{1 \leq i \leq n} (v^d_i \leftrightarrow y_i) 
\end{align*}

\subsection{Cardinality encoding}

This is the contribution of this paper to the SAT encoding of sorting networks. The previous encodings limit the number of layers to $d$. In order to perform a joint depth and size optimization we include additional clauses to limit the total number of comparators to $s$. 

Encodings of cardinality constraints into SAT have been thoroughly studied over the last few years. Interestingly, a good solution for our case is to use cardinality encodings based on sorting networks.

In a binary sorting network that takes input variables $(x_1 \ldots x_n)$ and returns the sorted version in decreasing order $(y_1 \ldots y_n)$ the output variable $y_s$ becomes true if and only if there are at least $s$ true input variables. Therefore, to express $x_1 + \ldots + x_n \leq s$ it suffices to add a unit clause $\neg y_{s+1}$. Standard cardinality encodings based on sorting networks requires $O(n \log^2 n)$ clauses and variables. However, the selected cardinality encoding proposed by Ab\'io et al. \cite{abio2013parametric} reduces this to $O(n \log^2 k)$ and enforces arc-consistency, which represents a significant optimization in our case.

A detailed enumeration of all the clauses and variables of the selected cardinality encoding is out of the scope of this paper. We assume here that we are given the cardinality variable $c_{s+1}$ and the corresponding $K$ cardinality clauses $u_k$, with $1 \leq k \leq K$

\begin{align*}
(c_{s+1}; u_1, \ldots, u_{K}) = Card_{s+1}(C^d_n)
\end{align*} 

where $c_{s+1}$ is false and all the cardinality clauses are satisfied if and only if  there are $s$ or less comparator variables that are true.

\begin{align*}
\less_{s+1}(C^d_n) = \neg c_{s+1}  \wedge  \bigwedge_{1 \leq k \leq K} u_k
\end{align*}

\subsection{Basic encoding}

A sorting network for $n$ channels on $d$ layers with $s$ or less comparators exists if and only if the following constraint is satisfiable.
\begin{equation}\label{eq:sat}
\varphi(n,d,s) = \valid(C^d_n) \wedge \less_{s+1}(C^d_n) \wedge \bigwedge_{\bar x \in \{0,1\}^n} \sorts(C^d_n, \bar x)
\end{equation}

\subsection{Additional encodings}
The basic sorting network encoding can be improved with additional constraints that restrict the search space or help to find conflicts quickly, as well as other optimizations described in \cite{codish2016sorting}. In this subsection we just enumerate the additional constraints considered in our SAT encoding. The reader is refereed to \cite{codish2016sorting} for a detailed justification.

\begin{description}
\item[Redundant \textit{sorts} clauses.]
The following encoding adds specific redundant $\sorts$ clauses that allows for more propagations, thus conflicts can be found earlier

\begin{align*}
\OD^k_{i, j} & \leftrightarrow \bigvee_{i < \ell \leq j} g^k_{i,\ell} &
\ND^k_{i, j} & \leftrightarrow \neg \OD^k_{i, j} \\
\OU^k_{i, j} &\leftrightarrow \bigvee_{i \leq \ell < j} g^k_{\ell, j} & 
\NU^k_{i, j} &\leftrightarrow \neg \OU^k_{i, j}
\end{align*}

Given an input $\vec x = (0, 0, \ldots, 0, x_t, x_{t+1}, \ldots, x_{t+r-1}, 1, 1, \ldots, 1) $, for all $t \leq i \leq t+r-1$ and at each layer $k$, we add the following constraints to the definition of $\sorts$

\begin{align*}
\bigwedge_{1 \leq k \leq d } v^{k-1}_i \wedge \ND^k_{i, t+r-1} &\rightarrow v^{k}_i \\
\bigwedge_{1 \leq k \leq d } \neg v^{k-1}_i \wedge \NU^k_{t, i} &\rightarrow \neg v^{k}_i  
\end{align*}
\end{description}

Other additional constraints of interest in our case follows the necessary conditions for the last layers \cite{codish2016sorting}. However, we can not use constraints that may force redundant comparators. In our optimal networks the last layer can have adjacent unused channels. 

\begin{description}
\item[Non-redundant comparators in the last layer connect adjacent channels.]

\[\varphi_1 = \sset{\neg g^d_{i,j}}{1\leq i,i+1<j\leq n}\]

\item[No comparator in the penultimate layer connect two channels that are more than 3 channels apart.]

\[\varphi_2 = \sset{\neg g^{d-1}_{i,j}}{1\leq i,i+3<j\leq n}\]

\item[A comparator $(i,i+2)$ or $(i,i+2)$ on the penultimate layer has implications in the last layer.]

\[\varphi_3 = \sset{g^{d-1}_{i,i+3} \to g^d_{i,i+1}\right) \wedge \left(g^{d-1}_{i,i+3} \to g^d_{i+2,i+3}}{1\leq i\leq n-3}\]
\[\varphi_4 = \sset{g^{d-1}_{i,i+2}\to g^d_{i,i+1}\vee g^d_{i+1,i+2}}{1\leq i\leq n-2}\]

\end{description}

And we also included the additional optimizations from \cite{bundala2014optimal}

\begin{description}
\item[No redundant comparators.]

\[
\sigma_1 = \hspace{-5mm}
\bigwedge_{\scriptsize\begin{array}{c}
	1 \leq k < d\\
	1 \leq i<j \leq n
	\end{array}}\hspace{-3mm}
\neg g^k_{i,j} \vee \neg g^{k+1}_{i,j}
\]
\item[Eager comparator placement.]

\[
\sigma_2 = \hspace{-5mm}
\bigwedge_{\scriptsize\begin{array}{c}
	1 < k \leq d\\
	1 \leq i<j \leq n
	\end{array}}\hspace{-3mm}
g^k_{i,j} \rightarrow \used^{k-1}_{i}(C^d_n) \vee \used^{k-1}_{j}(C^d_n)
\]

\item[All adjacent comparators.]

\[
\sigma_3 = \hspace{-5mm}
\bigwedge_{\scriptsize\begin{array}{c}
	1 \leq i < n
	\end{array}}\hspace{-3mm}
\left( g^1_{i,i+1} \vee g^2_{i,i+1} \vee\cdots\vee g^d_{i,i+1}
\right)
\]

\item[Only unsorted inputs.]
We can remove sort constraints $\sorts(C^d_n, x)$ for already sorted inputs from the basic encoding. Sorted inputs always remain unchanged.

\end{description}

Another key tool to obtain a tractable SAT encoding is to consider a fixed prefix. If we fix the first layers of the networks we not only reduce the search space in term of free comparators but also the number of sort constraints $\sorts(C^d_n, x)$ and validity clauses. In the sorting encodings we have to consider only the rest of the network and the remaining unsorted sequences at the output of fixed prefix.

We study the generation of a complete set of prefixes for our optimality results in the following section.

\section{Symbolic representation of two-layer prefixes}
\label{sec:symbolic}

\begin{figure}[h]
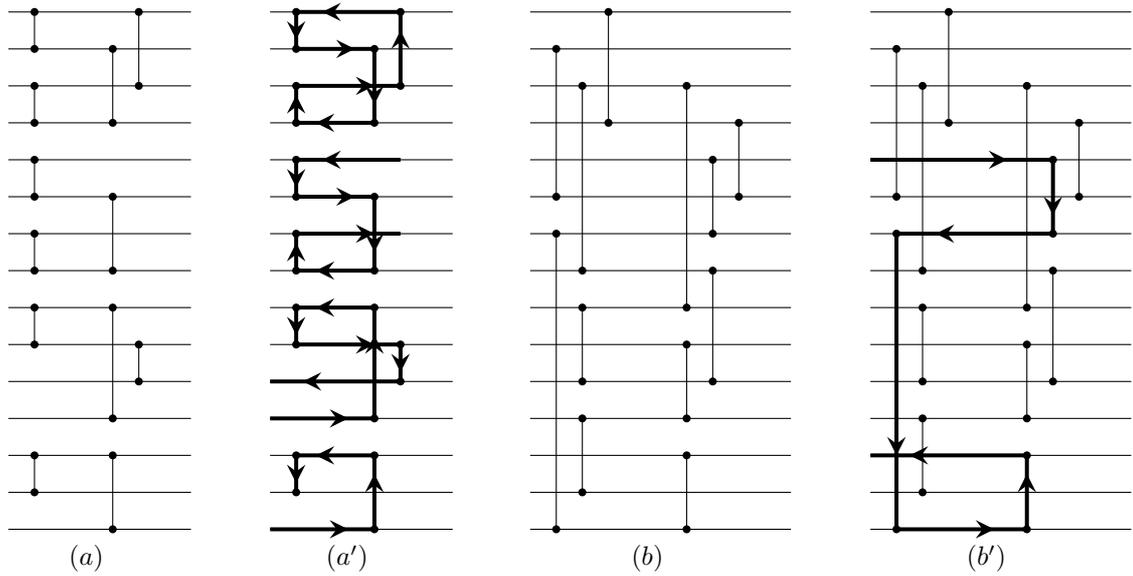

	\centering
	\begin{tabular}{cccc}
		\begin{sortingnetwork}{15}{0.7}
			\nodeconnectionref{ {1, 2}, {5, 6}, {7, 8}, {9, 10}, {11, 12}, {13, 14}}
			\addtocounter{sncolumncounter}{2}
			\nodeconnectionref{ {0, 2}, {3, 6}, {7, 9}, {11, 13}}
			\nodeconnectionref{ {4, 5}, {12, 14}}
		\end{sortingnetwork}
		&
		\begin{sortingnetwork}{15}{0.7}
			\nodeconnectionref{ {1, 2}, {5, 6}, {7, 8}, {9, 10}, {11, 12}, {13, 14}}
			\addtocounter{sncolumncounter}{2}
			\nodeconnectionref{ {0, 2}, {3, 6}, {7, 9}, {11, 13}}
			\nodeconnectionref{ {4, 5}, {12, 14}}
			
			\addarrow{-1}131
			\addarrow3133
			\addarrow3303
			\addarrow0302
			
			\addarrow{-1}434
			\addarrow3437
			\addarrow3707
			\addarrow0706
			\addarrow0646
			\addarrow4645
			\addarrow45{-1}5
			
			\addarrow4{11}0{11}
			\addarrow0{11}0{10}
			\addarrow0{10}3{10}
			\addarrow3{10}3{8}
			\addarrow3{8}0{8}
			\addarrow0{8}0{9}
			\addarrow0{9}4{9}
			
			\addarrow4{15}0{15}
			\addarrow0{15}0{14}
			\addarrow0{14}3{14}
			\addarrow3{14}3{12}
			\addarrow3{12}0{12}
			\addarrow0{12}0{13}
			\addarrow0{13}4{13}
			\addarrow4{13}4{15}
			
		\end{sortingnetwork}
		&
		\begin{sortingnetwork}{15}{0.7}
			\nodeconnectionref{ {0, 8}, {9, 13}}
			\nodeconnectionref{ {1, 3}, {4, 6}, {7, 12}}
			\nodeconnectionref{ {11, 14}}
			\addtocounter{sncolumncounter}{2}
			\nodeconnectionref{ {0, 2}, {3, 5}, {6, 12}}
			\nodeconnectionref{ {4, 7}, {8, 10}}
			\nodeconnectionref{ {9, 11}}
		\end{sortingnetwork}
		&
		\begin{sortingnetwork}{15}{0.7}
			\nodeconnectionref{ {0, 8}, {9, 13}}
			\nodeconnectionref{ {1, 3}, {4, 6}, {7, 12}}
			\nodeconnectionref{ {11, 14}}
			\addtocounter{sncolumncounter}{2}
			\nodeconnectionref{ {0, 2}, {3, 5}, {6, 12}}
			\nodeconnectionref{ {4, 7}, {8, 10}}
			\nodeconnectionref{ {9, 11}}
			
			\addarrow{-1}{11}6{11}
			\addarrow6{11}6{9}
			\addarrow6{9}0{9}
			\addarrow0{9}0{1}
			\addarrow0{1}5{1}
			\addarrow5{1}5{3}
			\addarrow5{3}{-1}{3}
			
		\end{sortingnetwork}
		\\
		$(a)$&$(a')$&$(b)$&$(b')$
	\end{tabular}
	\caption{The two-layer networks $(a)$ and $(b)$ are equivalent up to permutation. The symbolic representation of both networks is $\sent{(012,0120,1221,1221c)}$. The 4 words of this sentence are obtained from the 4 maximal paths represented in $(a')$ for each connected component. The path of $(b)$ corresponding to the word $\sent{0120}$ is also depicted in $(b')$.}
	\label{ex:paths}
\end{figure}

Bundala et al. \cite{bundala2014optimal-depth} studied the characterization and symbolic representation of equivalent two-layer prefixes up to permutation for the optimal-depth search problem. The proposed symbolic representation is based on the observation that two-layer networks are fully characterized by the maximal-length simple path of each group of connected comparators. Figure \ref{ex:paths} shows two equivalent two-layer networks $(a)$ and $(b)$, all the maximal paths of network $(a')$, and one maximal path of network $(b')$.

Using this observation and additional symmetry properties of maximal-length paths, Bundala et al. developed an efficient method to generate a \emph{complete set of prefixes} for the optimal-depth sorting network problem. The generation algorithm considered only networks with a maximal first layer (with $\left\lfloor\frac{n}{2}\right\rfloor$ comparators) and \emph{saturated} prefixes (\cite{bundala2014optimal}, Definition 7).  

For the joint depth-size optimization problem we can not keep those restrictions. We need to address the general case of isomorphic two-layer comparator networks with any number of comparators in the first and second layer.

The proposed symbolic generation of a complete set of prefixes is an extension of the generation algorithm of Bundala et al. \cite{bundala2014optimal} including non-maximal first layers. We follow a similar terminology and definitions.

The smaller (resp.\ larger) channel in some comparator of the first layer is called a \emph{min-channel} (respectively, a \emph{max-channel}) and an unused channel in the first layer will be called a \emph{free} channels. In our case, we can have more than one free channel.

\begin{definition}\label{def:path}
	(Bundala et al. \cite{bundala2014optimal}) A \emph{path} in a two-layer network $C$ is a sequence
	$\tuple{p_1p_2\ldots p_k}$ of distinct channels such that each pair of consecutive channels is connected by a comparator in~$C$.
	
	The \emph{word} corresponding to $\tuple{p_1p_2\ldots p_k}$ is $\tuple{w_1w_2\ldots w_k}$, where:
	\[w_i=\begin{cases}
	\sent{0} & \mbox{if $p_i$ is a free channel} \\
	\sent{1} & \mbox{if $p_i$ is a min-channel} \\
	\sent{2} & \mbox{if $p_i$ is a max-channel}
	\end{cases}\]
\end{definition}

A path is maximal if it is a simple path (with no repeated nodes) that cannot be extended (in either direction). A network is connected if its graph representation is connected. In general, we can have connected networks with up to two free channels, so we need to consider one additional type of word that we will name \emph{Tail-word}.

\begin{figure}
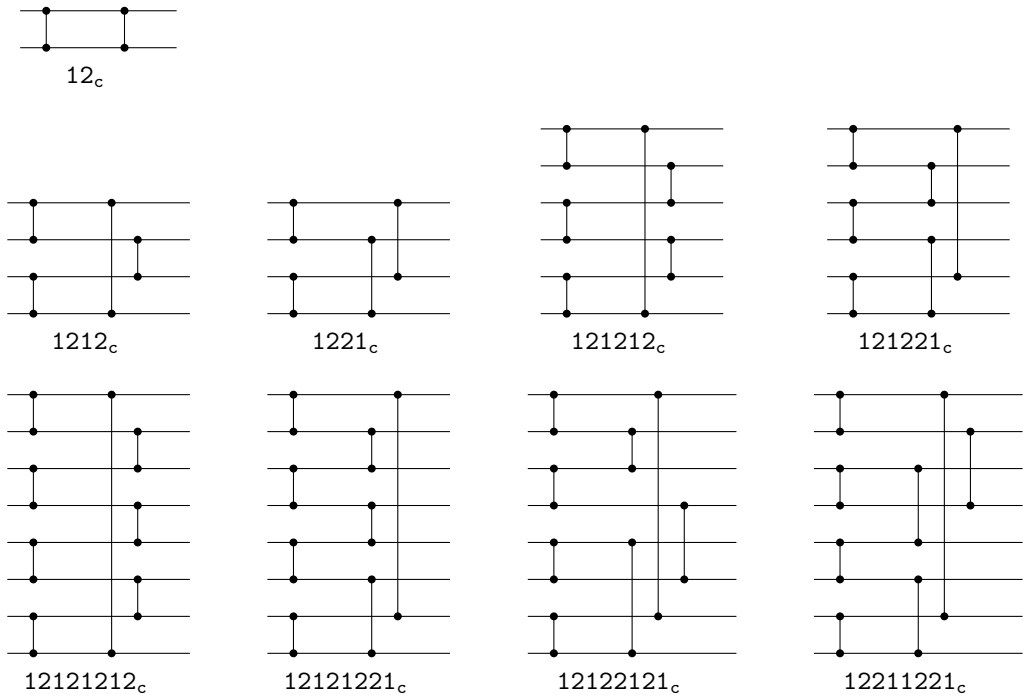

	\begin{tabular}{cccc}
		\begin{sortingnetwork}{2}{0.7}
			\nodeconnectionref{ {0, 1}}
			\addtocounter{sncolumncounter}{2}
			\nodeconnectionref{ {0, 1}}
		\end{sortingnetwork}
		\\
		$\sent{12_c}$    \\
		\\
		\begin{sortingnetwork}{4}{0.7}
			\nodeconnectionref{ {0, 1}, {2, 3}}
			\addtocounter{sncolumncounter}{2}
			\nodeconnectionref{ {0, 3}}
			\nodeconnectionref{ {1, 2}}
		\end{sortingnetwork}
		&
		\begin{sortingnetwork}{4}{0.7}
			\nodeconnectionref{ {0, 1}, {2, 3}}
			\addtocounter{sncolumncounter}{2}
			\nodeconnectionref{ {0, 2}}
			\nodeconnectionref{ {1, 3}}
		\end{sortingnetwork}
		&
		\begin{sortingnetwork}{6}{0.7}
			\nodeconnectionref{ {0, 1}, {2, 3}, {4, 5}}
			\addtocounter{sncolumncounter}{2}
			\nodeconnectionref{ {0, 5}}
			\nodeconnectionref{ {1, 2}, {3, 4}}
		\end{sortingnetwork}
		&
		\begin{sortingnetwork}{6}{0.7}
			\nodeconnectionref{ {0, 1}, {2, 3}, {4, 5}}
			\addtocounter{sncolumncounter}{2}
			\nodeconnectionref{ {0, 2}, {3, 4}}
			\nodeconnectionref{ {1, 5}}
		\end{sortingnetwork}
		\\
		$\sent{1212_c}$ & $\sent{1221_c}$ & $\sent{121212_c}$ & $\sent{121221_c}$    \\
		\\
		\begin{sortingnetwork}{8}{0.7}
			\nodeconnectionref{ {0, 1}, {2, 3}, {4, 5}, {6, 7}}
			\addtocounter{sncolumncounter}{2}
			\nodeconnectionref{ {0, 7}}
			\nodeconnectionref{ {1, 2}, {3, 4}, {5, 6}}
		\end{sortingnetwork}
		&
		\begin{sortingnetwork}{8}{0.7}
			\nodeconnectionref{ {0, 1}, {2, 3}, {4, 5}, {6, 7}}
			\addtocounter{sncolumncounter}{2}
			\nodeconnectionref{ {0, 2}, {3, 4}, {5, 6}}
			\nodeconnectionref{ {1, 7}}
		\end{sortingnetwork}
		&
		\begin{sortingnetwork}{8}{0.7}
			\nodeconnectionref{ {0, 1}, {2, 3}, {4, 5}, {6, 7}}
			\addtocounter{sncolumncounter}{2}
			\nodeconnectionref{ {0, 3}, {5, 6}}
			\nodeconnectionref{ {1, 7}}
			\nodeconnectionref{ {2, 4}}
		\end{sortingnetwork}
		&
		\begin{sortingnetwork}{8}{0.7}
			\nodeconnectionref{ {0, 1}, {2, 3}, {4, 5}, {6, 7}}
			\addtocounter{sncolumncounter}{2}
			\nodeconnectionref{ {0, 2}, {3, 5}}
			\nodeconnectionref{ {1, 7}}
			\nodeconnectionref{ {4, 6}}
		\end{sortingnetwork}

		\\
		$\sent{12121212_c}$ &$\sent{12121221_c}$ &$\sent{12122121_c}$ &$\sent{12211221_c}$    \\
	\end{tabular}
	\caption{Complete set of Cycle-words on $n\le8$ channels.}
	\label{ex:cycle}
\end{figure}

\begin{definition}
	\label{defn:word}
	Let $C$ be a connected two-layer network on $n$ channels. We will classify $C$ and its corresponding \emph{word} based on the number of unused channels in the first and second layer. In a connected two-layer network we can have only four different kinds of words. 

	\begin{description}
		\item[Head-word.] If $n$ is odd, then $\word(C)$ is the word corresponding to the maximal path in $C$ starting with the (unique) free channel. The number of Head-words on $n$ channels is $2^{(n-1)/2}$. Figure \ref{ex:head} show the complete set of Head-words on $n\le5$ channels.
		
		\item[Stick-word.] If $n$ is even and $C$ has two channels not used in layer~$2$, then $\word(C)$ is the lexicographically smallest of the words corresponding to the two maximal paths in $C$ starting with one of these unused channels (which are reverse to one another). The number of Stick-words on $(2,4,6,8,10,12,14,16,...)$ channels is $(1,3,4,10,16,36,64,136,...)$ respectively (OEIS A051437)\footnote{\href{https://oeis.org/A051437}{The On-Line Encyclopedia of Integer Sequences, Sequence A051437}}. Figure \ref{ex:stick} show the complete set of Stick-words on $n\le6$ channels.

		\item[Cycle-word.] If $n$ is even and all channels are used by a comparator in layer $2$, then $\word(C)$ is the lexicographically smallest word corresponding to a maximal path in $C$ that begins with two channels connected in layer~$1$. The number of Cycle-words on $(2,4,6,8,10,12,14,16,...)$ channels is $(1,2,2,4,4,9,10,22,...)$ respectively (OEIS A053656)\footnote{\href{https://oeis.org/A053656}{The On-Line Encyclopedia of Integer Sequences, Sequence A053656}}. Figure \ref{ex:cycle} show the complete set of Stick-words on $n\le8$ channels.

		\item[Tail-word.] If $n$ is even and $C$ has two free channels then $\word(C)$ is the lexicographically smallest of the words corresponding to the two maximal paths in $C$ starting with each of the two free channels (which are reverse to one another). Each two-layer network represented by a Tail-word is equal to a Stick-word network with two additional free channels and two comparators connecting each of the unused channels in the second layer with these free channels. The number of Head-words on $n$ channels is equal to the number of Tail-words in $n-2$ channels. Figure \ref{ex:tail} show the complete set of Tail-words on $n\le8$ channels.

	\end{description}
\end{definition}

\begin{figure}
	\begin{tabular}{cccc}
		\begin{sortingnetwork}{1}{0.7}
			\addtocounter{sncolumncounter}{5}
		\end{sortingnetwork}
		&
		\begin{sortingnetwork}{3}{0.7}
			\nodeconnectionref{ {1, 2}}
			\addtocounter{sncolumncounter}{2}
			\nodeconnectionref{ {0, 2}}
		\end{sortingnetwork}
		&
		\begin{sortingnetwork}{3}{0.7}
			\nodeconnectionref{ {1, 2}}
			\addtocounter{sncolumncounter}{2}
			\nodeconnectionref{ {0, 1}}
		\end{sortingnetwork}
		\\
		$\sent{0}$ & $\sent{012}$ & $\sent{021}$
		\\
		\\
		\begin{sortingnetwork}{5}{0.7}
			\nodeconnectionref{ {1, 2}, {3, 4}}
			\addtocounter{sncolumncounter}{2}
			\nodeconnectionref{ {0, 4}}
			\nodeconnectionref{ {2, 3}}
		\end{sortingnetwork}
		&
		\begin{sortingnetwork}{5}{0.7}
			\nodeconnectionref{ {1, 2}, {3, 4}}
			\addtocounter{sncolumncounter}{2}
			\nodeconnectionref{ {0, 4}}
			\nodeconnectionref{ {1, 3}}
		\end{sortingnetwork}
		&
		\begin{sortingnetwork}{5}{0.7}
			\nodeconnectionref{ {1, 2}, {3, 4}}
			\addtocounter{sncolumncounter}{2}
			\nodeconnectionref{ {0, 3}}
			\nodeconnectionref{ {2, 4}}
		\end{sortingnetwork}
		&
		\begin{sortingnetwork}{5}{0.7}
			\nodeconnectionref{ {1, 2}, {3, 4}}
			\addtocounter{sncolumncounter}{2}
			\nodeconnectionref{ {0, 3}}
			\nodeconnectionref{ {1, 4}}
		\end{sortingnetwork}		
		\\
		$\sent{01212}$ & $\sent{01221}$ & $\sent{02112}$  & $\sent{02121}$
	\end{tabular}
	\caption{Complete set of Head-words on $n\le5$ channels.}
	\label{ex:head}
\end{figure}

\begin{figure}
	\begin{tabular}{cccc}
		\begin{sortingnetwork}{2}{0.7}
			\nodeconnectionref{ {0, 1}}
			\addtocounter{sncolumncounter}{3}
		\end{sortingnetwork}
		\\
		$\sent{12}$    \\
		\\
		\begin{sortingnetwork}{4}{0.7}
			\nodeconnectionref{ {0, 1}, {2, 3}}
			\addtocounter{sncolumncounter}{2}
			\nodeconnectionref{ {1, 2}}
		\end{sortingnetwork}
		
		&
		\begin{sortingnetwork}{4}{0.7}
			\nodeconnectionref{ {0, 1}, {2, 3}}
			\addtocounter{sncolumncounter}{2}
			\nodeconnectionref{ {0, 2}}
		\end{sortingnetwork}
		
		&
		\begin{sortingnetwork}{4}{0.7}
			\nodeconnectionref{ {0, 1}, {2, 3}}
			\addtocounter{sncolumncounter}{2}
			\nodeconnectionref{ {1, 3}}
		\end{sortingnetwork}
		
		\\
		$\sent{1212}$ &$\sent{1221}$ &$\sent{2112}$    \\
		\\
		\begin{sortingnetwork}{6}{0.7}
			\nodeconnectionref{ {0, 1}, {2, 3}, {4, 5}}
			\addtocounter{sncolumncounter}{2}
			\nodeconnectionref{ {1, 2}, {3, 4}}
		\end{sortingnetwork}
		
		&
		\begin{sortingnetwork}{6}{0.7}
			\nodeconnectionref{ {0, 1}, {2, 3}, {4, 5}}
			\addtocounter{sncolumncounter}{2}
			\nodeconnectionref{ {0, 2}, {3, 4}}
		\end{sortingnetwork}
		
		&
		\begin{sortingnetwork}{6}{0.7}
			\nodeconnectionref{ {0, 1}, {2, 3}, {4, 5}}
			\addtocounter{sncolumncounter}{2}
			\nodeconnectionref{ {1, 3}}
			\nodeconnectionref{ {2, 4}}
		\end{sortingnetwork}
		
		&
		\begin{sortingnetwork}{6}{0.7}
			\nodeconnectionref{ {0, 1}, {2, 3}, {4, 5}}
			\addtocounter{sncolumncounter}{2}
			\nodeconnectionref{ {1, 2}, {3, 5}}
		\end{sortingnetwork}
		
		\\
		$\sent{121212}$ &$\sent{121221}$ &$\sent{122112}$ &$\sent{211212}$
		\\
	\end{tabular}
	\caption{Complete set of Stick-words on $n\le6$ channels.}
	\label{ex:stick}
\end{figure}

\begin{figure}
	\begin{tabular}{cccc}
		\begin{sortingnetwork}{4}{0.7}
			\nodeconnectionref{ {2, 3}}
			\addtocounter{sncolumncounter}{2}
			\nodeconnectionref{ {0, 3}}
			\nodeconnectionref{ {1, 2}}
		\end{sortingnetwork}
		
		\\
		$\sent{0120}$    \\
		\\
		\begin{sortingnetwork}{6}{0.7}
			\nodeconnectionref{ {2, 3}, {4, 5}}
			\addtocounter{sncolumncounter}{2}
			\nodeconnectionref{ {0, 5}}
			\nodeconnectionref{ {1, 2}, {3, 4}}
		\end{sortingnetwork}
		
		&
		\begin{sortingnetwork}{6}{0.7}
			\nodeconnectionref{ {2, 3}, {4, 5}}
			\addtocounter{sncolumncounter}{2}
			\nodeconnectionref{ {0, 5}}
			\nodeconnectionref{ {1, 3}}
			\nodeconnectionref{ {2, 4}}
		\end{sortingnetwork}
		
		&
		\begin{sortingnetwork}{6}{0.7}
			\nodeconnectionref{ {2, 3}, {4, 5}}
			\addtocounter{sncolumncounter}{2}
			\nodeconnectionref{ {0, 4}}
			\nodeconnectionref{ {1, 2}, {3, 5}}
		\end{sortingnetwork}
		
		\\
		$\sent{012120}$ &$\sent{012210}$ &$\sent{021120}$    \\
		\\
		\begin{sortingnetwork}{8}{0.7}
			\nodeconnectionref{ {2, 3}, {4, 5}, {6, 7}}
			\addtocounter{sncolumncounter}{2}
			\nodeconnectionref{ {0, 7}}
			\nodeconnectionref{ {1, 2}, {3, 4}, {5, 6}}
		\end{sortingnetwork}
		
		&
		\begin{sortingnetwork}{8}{0.7}
			\nodeconnectionref{ {2, 3}, {4, 5}, {6, 7}}
			\addtocounter{sncolumncounter}{2}
			\nodeconnectionref{ {0, 7}}
			\nodeconnectionref{ {1, 3}, {5, 6}}
			\nodeconnectionref{ {2, 4}}
		\end{sortingnetwork}
		
		&
		\begin{sortingnetwork}{8}{0.7}
			\nodeconnectionref{ {2, 3}, {4, 5}, {6, 7}}
			\addtocounter{sncolumncounter}{2}
			\nodeconnectionref{ {0, 7}}
			\nodeconnectionref{ {1, 2}, {3, 5}}
			\nodeconnectionref{ {4, 6}}
		\end{sortingnetwork}
		
		&
		\begin{sortingnetwork}{8}{0.7}
			\nodeconnectionref{ {2, 3}, {4, 5}, {6, 7}}
			\addtocounter{sncolumncounter}{2}
			\nodeconnectionref{ {0, 6}}
			\nodeconnectionref{ {1, 2}, {3, 4}, {5, 7}}
		\end{sortingnetwork}
		\\
		$\sent{01212120}$ &$\sent{01212210}$ &$\sent{01221120}$ &$\sent{02112120}$    \\
	\end{tabular}
	\caption{Complete set of Tail-words on $n\le8$ channels.}
	\label{ex:tail}
\end{figure}

The set of all possible words (not necessarily minimal with respect to lexicographic ordering) can be described by the following BNF-style grammar\footnote{ We do not include the Tail-word description of a network with a single comparator in the second layer, because it is already covered by the equivalent single comparator in the first layer, Stick-word $\sent{12}$}.

\nopagebreak

\begin{align}
\label{eq:word}
\mathsf{Word} &::= \mathsf{Head} \mid \mathsf{Tail} \mid \mathsf{Stick} \mid \mathsf{Cycle}\\
\nonumber
\mathsf{Head} &::= \sent{0}(\sent{12} \mid \sent{21})^\ast
& \mathsf{Stick} &::= (\sent{12} \mid \sent{21})^+ \\
\nonumber
\mathsf{Tail} &::= \sent{0}(\sent{12} \mid \sent{21})^+{0}
& \mathsf{Cycle} &::= \sent{12}(\sent{12} \mid \sent{21})^+
\end{align}
To avoid ambiguity with Stick-words, we annotate Cycle-words with a \sent c\ tag.

\begin{definition}
	\label{defn:sentence}
	A two-layer comparator network $C$ is represented by the multi-set $word(C)$ containing $w'=word(C')$ for each connected component $C'$ of $C$. The set is denoted by the \emph{sentence} $w_1;w_2;\ldots;w_k$ where the words are in lexicographic order. 
\end{definition}

Figure~\ref{ex:paths} illustrates the case of a two-layer network with 4 connected components, one of each type.

\begin{definition}
	\label{defn:net}
	Let $w$ be a word in the language of Equation~\eqref{eq:word}, and $n=|w|$.
	The two-layer network $\net(w)$ has a first layer with $m$ comparators of the form $(2i-1,2i)$, with $1 \le i \le\ m$, where:
	\[m=\begin{cases}
	\frac n2 & \mbox{if $w$ is a Stick-word or a Cycle-word} \\
	\frac{n-1}2 & \mbox{if $w$ is a Cycle-word} \\
	\frac{n-2}2 & \mbox{if $w$ is a Tail-word}
	\end{cases}\]
	The second layer is then defined as follows.
	\begin{enumerate}
		\item If $w$ is a Stick-word or a Cycle-word, ignore the first character; then, for $k=0,\ldots,\left\lfloor\frac n2\right\rfloor-1$, take the next two characters $xy$ of $w$ and add a second-layer comparator between channels $2k+x$ and $2(k+1)+y$. Ignore the last character; if $w$ is a Cycle-word, connect the two remaining channels at the end. (Figure \ref{ex:stick} and \ref{ex:cycle})
		\item If $w$ is a Head-word, proceed as above but start by connecting the free channel to the channel indicated by the second character. (Figure \ref{ex:head})
		\item If $w$ is a Tail-word, ignore the zeros and proceed as for a Stick-word. Then connect the channel indicated by second character with the second free channel, and the remaining channel indicated by the penultimate character with the first free channel. (Figure \ref{ex:tail})
	\end{enumerate}
\end{definition}

To generate a network from a sentence, we generate the network of each word and compose them bottom-up as illustrated in Figure~\ref{ex:paths}$(a)$ for the sentence $\sent{(012,0120,1221,1221c)}$.

\begin{figure}
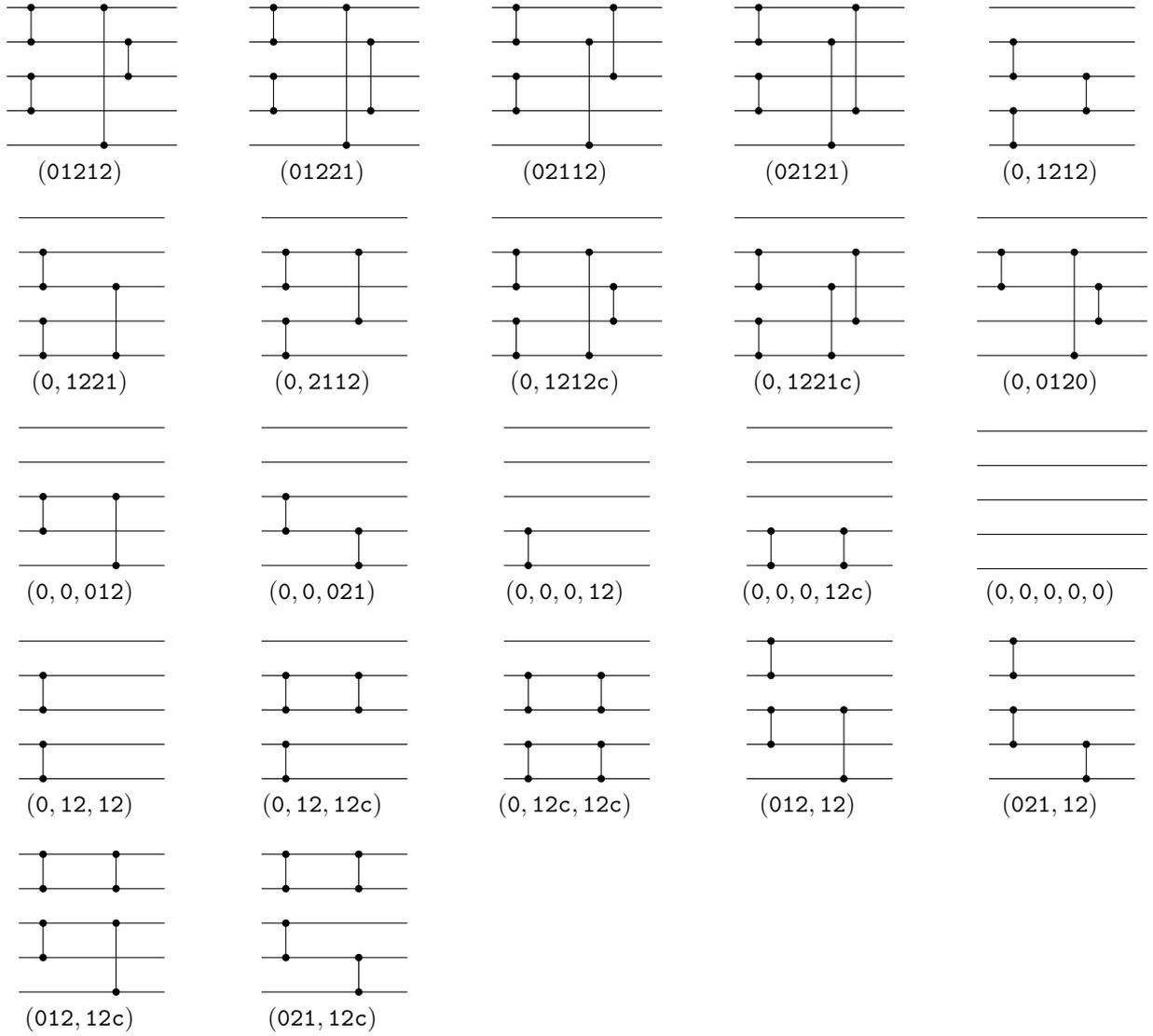

	\begin{tabular}{ccccc}
\begin{sortingnetwork}{5}{0.7}
	\nodeconnectionref{ {1, 2}, {3, 4}}
	\addtocounter{sncolumncounter}{2}
	\nodeconnectionref{ {0, 4}}
	\nodeconnectionref{ {2, 3}}
\end{sortingnetwork}
&
\begin{sortingnetwork}{5}{0.7}
	\nodeconnectionref{ {1, 2}, {3, 4}}
	\addtocounter{sncolumncounter}{2}
	\nodeconnectionref{ {0, 4}}
	\nodeconnectionref{ {1, 3}}
\end{sortingnetwork}
&
\begin{sortingnetwork}{5}{0.7}
	\nodeconnectionref{ {1, 2}, {3, 4}}
	\addtocounter{sncolumncounter}{2}
	\nodeconnectionref{ {0, 3}}
	\nodeconnectionref{ {2, 4}}
\end{sortingnetwork}
&
\begin{sortingnetwork}{5}{0.7}
	\nodeconnectionref{ {1, 2}, {3, 4}}
	\addtocounter{sncolumncounter}{2}
	\nodeconnectionref{ {0, 3}}
	\nodeconnectionref{ {1, 4}}
\end{sortingnetwork}
&
\begin{sortingnetwork}{5}{0.7}
	\nodeconnectionref{ {0, 1}, {2, 3}}
	\addtocounter{sncolumncounter}{2}
	\nodeconnectionref{ {1, 2}}
\end{sortingnetwork}
\\
$\sent{(01212)}$ & $\sent{(01221)}$ & $\sent{(02112)}$ & $\sent{(02121)}$ & $\sent{(0,1212)}$
\\
\\
\begin{sortingnetwork}{5}{0.7}
	\nodeconnectionref{ {0, 1}, {2, 3}}
	\addtocounter{sncolumncounter}{2}
	\nodeconnectionref{ {0, 2}}
\end{sortingnetwork}
&
\begin{sortingnetwork}{5}{0.7}
	\nodeconnectionref{ {0, 1}, {2, 3}}
	\addtocounter{sncolumncounter}{2}
	\nodeconnectionref{ {1, 3}}
\end{sortingnetwork}
&
\begin{sortingnetwork}{5}{0.7}
	\nodeconnectionref{ {0, 1}, {2, 3}}
	\addtocounter{sncolumncounter}{2}
	\nodeconnectionref{ {0, 3}}
	\nodeconnectionref{ {1, 2}}
\end{sortingnetwork}
&
\begin{sortingnetwork}{5}{0.7}
	\nodeconnectionref{ {0, 1}, {2, 3}}
	\addtocounter{sncolumncounter}{2}
	\nodeconnectionref{ {0, 2}}
	\nodeconnectionref{ {1, 3}}
\end{sortingnetwork}
&
\begin{sortingnetwork}{5}{0.7}
	\nodeconnectionref{ {2, 3}}
	\addtocounter{sncolumncounter}{2}
	\nodeconnectionref{ {0, 3}}
	\nodeconnectionref{ {1, 2}}
\end{sortingnetwork}
\\
$\sent{(0,1221)}$ & $\sent{(0,2112)}$ & $\sent{(0,1212c)}$ & $\sent{(0,1221c)}$ & $\sent{(0,0120)}$
\\
\\
\begin{sortingnetwork}{5}{0.7}
	\nodeconnectionref{ {1, 2}}
	\addtocounter{sncolumncounter}{2}
	\nodeconnectionref{ {0, 2}}
\end{sortingnetwork}
&
\begin{sortingnetwork}{5}{0.7}
	\nodeconnectionref{ {1, 2}}
	\addtocounter{sncolumncounter}{2}
	\nodeconnectionref{ {0, 1}}
\end{sortingnetwork}
&
\begin{sortingnetwork}{5}{0.7}
	\nodeconnectionref{ {0, 1}}
	\addtocounter{sncolumncounter}{3}
\end{sortingnetwork}
&
\begin{sortingnetwork}{5}{0.7}
	\nodeconnectionref{ {0, 1}}
	\addtocounter{sncolumncounter}{2}
	\nodeconnectionref{ {0, 1}}
\end{sortingnetwork}
&
\begin{sortingnetwork}{5}{0.7}
	\addtocounter{sncolumncounter}{5}
\end{sortingnetwork}
\\
$\sent{(0,0,012)}$ & $\sent{(0,0,021)}$ & $\sent{(0,0,0,12)}$ & $\sent{(0,0,0,12c)}$ & $\sent{(0,0,0,0,0)}$
\\
\\
\begin{sortingnetwork}{5}{0.7}
	\nodeconnectionref{ {0, 1}, {2, 3}}
	\addtocounter{sncolumncounter}{3}
\end{sortingnetwork}
&
\begin{sortingnetwork}{5}{0.7}
	\nodeconnectionref{ {0, 1}, {2, 3}}
	\addtocounter{sncolumncounter}{2}
	\nodeconnectionref{ {2, 3}}
\end{sortingnetwork}
&
\begin{sortingnetwork}{5}{0.7}
	\nodeconnectionref{ {0, 1}, {2, 3}}
	\addtocounter{sncolumncounter}{2}
	\nodeconnectionref{ {0, 1}, {2, 3}}
\end{sortingnetwork}
&
\begin{sortingnetwork}{5}{0.7}
	\nodeconnectionref{ {1, 2}, {3, 4}}
	\addtocounter{sncolumncounter}{2}
	\nodeconnectionref{ {0, 2}}
\end{sortingnetwork}
&
\begin{sortingnetwork}{5}{0.7}
	\nodeconnectionref{ {1, 2}, {3, 4}}
	\addtocounter{sncolumncounter}{2}
	\nodeconnectionref{ {0, 1}}
\end{sortingnetwork}
\\
$\sent{(0,12,12)}$ & $\sent{(0,12,12c)}$ & $\sent{(0,12c,12c)}$ & $\sent{(012,12)}$ & $\sent{(021,12)}$
\\
\\
\begin{sortingnetwork}{5}{0.7}
	\nodeconnectionref{ {1, 2}, {3, 4}}
	\addtocounter{sncolumncounter}{2}
	\nodeconnectionref{ {0, 2}, {3, 4}}
\end{sortingnetwork}
&
\begin{sortingnetwork}{5}{0.7}
	\nodeconnectionref{ {1, 2}, {3, 4}}
	\addtocounter{sncolumncounter}{2}
	\nodeconnectionref{ {0, 1}, {3, 4}}
\end{sortingnetwork}
\\
$\sent{(012,12c)}$ & $\sent{(021,12c)}$
	\end{tabular}
    \caption{The complete set $R(H_5)$ of two-layer prefixes on 5 channels, including networks with redundant comparators:  $\sent{(0,0,0,12c)}$, $\sent{(0,12,12c)}$, $\sent{(0,12c,12c)}$, $\sent{(012,12c)}$, $\sent{(021,12c)}$, and networks without any comparator in the second layer: $\sent{(0,0,0,0,0)}$, $\sent{(0,0,0,12)}$, $\sent{(0,12,12)}$. Hence, $|R(H_5)| = 22$ and $|R(T_5)| = 22 - 5 - 3 = 14$.}
    \label{ex:prefixes}
\end{figure}

Figure \ref{ex:prefixes} depicts the 22 different 5-channel two-layer networks (up to permutation) and its corresponding sentences. This complete set includes the empty network $\sent{(0,0,0,0,0)}$ with no comparators, networks without any comparator in the second layer $\sent{(0,0,0,12)}$, $\sent{(0,12,12)}$, and networks with the word \sent{21c}, i.e., a redundant comparator.

\begin{lemma}
	\label{lem:equiv}
	\cite{bundala2014optimal}. Let $C$ and $C'$ be two-layer comparator networks on $n$ channels. Then $C \approx C'$ if and only if $\word(C)=\word(C')$.
\end{lemma}

We denote by $H_n^m$ the set of all possible $n$-channel two-layer network whose first layer has $m$ comparators of the form $(2i-1,2i)$, with $1 \le i \le\ m$ and $0 \le m \le \left\lfloor\frac n2\right\rfloor$. And by $H_n$ the union of the entire sequence.
\[
H_n = \bigcup_{m=0}^{\left\lfloor\frac n2\right\rfloor} H_n^m
\]
The set $H_n^m$ with $m=\left\lfloor\frac n2\right\rfloor$ is the set of networks with a fixed maximal first layer (denoted by $G_n$ in \cite{bundala2014optimal}). The set of representatives of the equivalence classes of $H_n$ and $G_n$ is denoted by $R(H_n)$ and $R(G_n)$ respectively.

For a given $n$ the set $R(H_n)$ can be generated from all multi-sets of valid words with a total of $n$ channels. Figure \ref{ex:prefixes} shows the complete $R(H_5)$ set. However, in the search for optimal networks we can remove prefixes with redundant comparators (word \sent{12c}), the empty network, and prefixes without any comparator in the second layer (with only the words \sent{0} and \sent{12} in its symbolic representation). We denote $R(T_n)$ the resulting reduced set of prefixes.

\begin{table}
	\[\begin{array}{c|r|r|r|r|r|r|r|r|r|r|r|r|r|r|r}
	n 
	& \multicolumn1{c|}{3}
	& \multicolumn1{c|}{4}
	& \multicolumn1{c|}{5}
	& \multicolumn1{c|}{6}
	& \multicolumn1{c|}{7}
	& \multicolumn1{c|}{8}
	& \multicolumn1{c|}{9}
	& \multicolumn1{c|}{10}
	& \multicolumn1{c|}{11}
	& \multicolumn1{c|}{12}
	& \multicolumn1{c|}{13}
	& \multicolumn1{c|}{14}
	& \multicolumn1{c|}{15}
    & \multicolumn1{c|}{16}
    & \multicolumn1{c}{17} \\ \hline
	|R(H_n)| & 5 & 14 & 22 & 50 & 84 & 178 & 300 & 588 & 1{,}004 & 1{,}900 & 3{,}234 & 5{,}904 & 10{,}054 & 17{,}959 & 30{,}435 \\ 
	|R(T_n)| & 2 & 8 & 14 & 32 & 58 & 123 & 211 & 404 & 698 & 1{,}305 & 2{,}223 & 3{,}996 & 6{,}812 & 12{,}046 & 20{,}372 \\ 
	|R(T'_n)| & 1 & 6 & 9 & 23 & 36 & 83 & 127 & 256 & 403 & 786 & 1{,}245 & 2{,}304 & 3{,}712 & 6{,}716 & 10{,}879 \\
	\hline
	|R(G_n)| & 4 & 8 & 16 & 20 & 52 & 61 & 165 & 152 & 482 & 414 & 1{,}378 & 1{,}024 & 3{,}780 & 2{,}627 & 10{,}187 \\ 
	|R(S_n)| & 2 & 2 & 6 & 6 & 14 & 15 & 37 & 27 & 88 & 70 & 212 & 136 & 494 & 323 & 1{,}149 \\ 
	|R(S'_n)| & 1 & 2 & 4 & 5 & 8 & 12 & 22 & 21 & 48 & 50 & 117 & 94 & 262 & 211 & 609 \\
	\hline
	\end{array}\]
	\[\begin{array}{c|r|r|r|r|r|r|r|r|r}
	n
	& \multicolumn1{c|}{18}
	& \multicolumn1{c|}{19}
	& \multicolumn1{c|}{20}
	& \multicolumn1{c|}{21}
	& \multicolumn1{c|}{22}
	& \multicolumn1{c|}{23}
	& \multicolumn1{c|}{24}
	& \multicolumn1{c|}{25}
	& \multicolumn1{c}{26} \\ \hline
	|R(H_n)| & 53{,}325 & 90{,}021 & 155{,}518 & 261{,}204 & 445{,}800 & 745{,}198 & 1{,}259{,}611 & 2{,}095{,}183 & 3{,}511{,}839 \\
	|R(T_n)| & 35{,}356 & 59{,}576 & 102{,}182 & 171{,}172 & 290{,}270 & 483{,}982 & 813{,}798 & 1{,}349{,}972 & 2{,}252{,}214 \\
	|R(T'_n)| & 19{,}191 & 31{,}301 & 54{,}352 & 88{,}847 & 152{,}011 & 248{,}867 & 421{,}233 & 689{,}320 & 1{,}155{,}520 \\
	\hline
	|R(G_n)| & 6{,}422 & 26{,}796 & 15{,}906 & 69{,}498 & 38{,}392 & 177{,}388 & 92{,}989 & 447{,}765 & 221{,}836 \\ 
	|R(S_n)| & 651 & 2{,}632 & 1{,}478 & 5{,}988 & 3{,}040 & 13{,}514 & 6{,}744 & 30{,}312 & 14{,}036 \\ 
	|R(S'_n)| & 411 & 1{,}367 & 894 & 3{,}098 & 1{,}787 & 6{,}920 & 3{,}848 & 15{,}469 & 7{,}830 \\
	\hline
	\end{array}\]
	
	\caption{Values of $|R(H_n)|$, $|R(T_n)|$, $|R(T'_n)|$, $|R(G_n)|$, $|R(S_n)|$ and $|R(S'_n)|$ for
		$n\leq 26$.}
	\label{tab:RHn}
\end{table}

\subsection{Reflections}

It is well-known that a reflection of a sorting network is also a sorting network \cite{knuth1998art, bundala2014optimal}. Formally, the reflection of comparator network $C$ is the network $C^R$ that replaces each comparator $(i,j)$ in $C$ with a comparator $(n-j+1,n-i+1)$ in $C^R$. Note that the reflection operation is not in general a permutation. Therefore, we can further reduce the number of prefixes in our complete set by removing those that are reflections of others. In the resulting symbolic representation of the complete set of two-layers prefixes, denoted by $R(T'_n)$, we keep the lexicographically smallest of the two sentences $\word(C)$ and $\word(C^R)$.

\begin{figure}
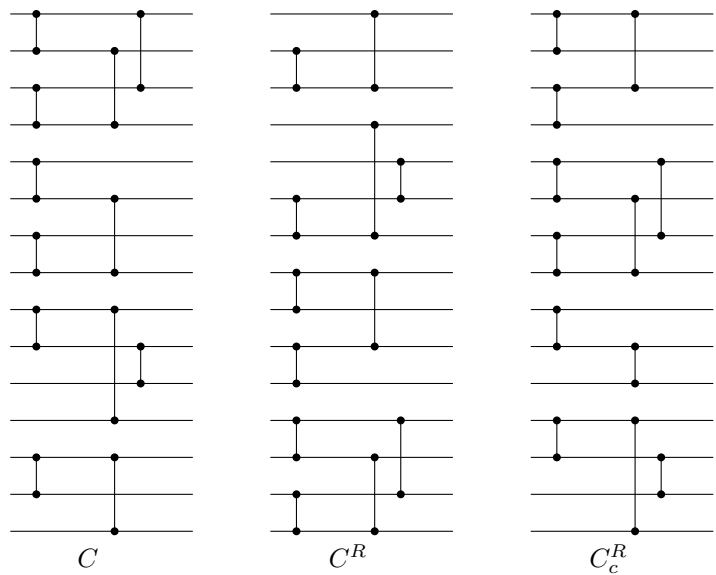

\centering
\begin{tabular}{ccc}
\begin{sortingnetwork}{15}{0.7}
	\nodeconnectionref{ {1, 2}, {5, 6}, {7, 8}, {9, 10}, {11, 12}, {13, 14}}
	\addtocounter{sncolumncounter}{2}
	\nodeconnectionref{ {0, 2}, {3, 6}, {7, 9}, {11, 13}}
	\nodeconnectionref{ {4, 5}, {12, 14}}
\end{sortingnetwork}
&
\begin{sortingnetwork}{15}{0.7}
	\nodeconnectionref{ {0, 1}, {2, 3}, {4, 5}, {6, 7}, {8, 9}, {12, 13}}
	\addtocounter{sncolumncounter}{2}
	\nodeconnectionref{ {0, 2}, {5, 7}, {8, 11}, {12, 14}}
	\nodeconnectionref{ {1, 3}, {9, 10}}
\end{sortingnetwork}
&
\begin{sortingnetwork}{15}{0.7}
	\nodeconnectionref{ {2, 3}, {5, 6}, {7, 8}, {9, 10}, {11, 12}, {13, 14}}
	\addtocounter{sncolumncounter}{2}
	\nodeconnectionref{ {0, 3}, {4, 5}, {7, 9}, {12, 14}}
	\nodeconnectionref{ {1, 2}, {8, 10}}
\end{sortingnetwork}
\\
$C$ & $C^R$ & $C^R_c$
	\end{tabular}
\caption{The network $C^R$ is the reflection of $C$, while $C^R_c = \net(\word(C^R))$ is the canonical permutation of network $C^R$. The symbolic representation of $C$ is $\sent{(012,0120,1221c,2112)}$ while the normalized symbolic representation of $C^R$ ($C^R_c$) is $\sent{(0120,021,1221c,2112)}$.}
\label{ex:reflection}
\end{figure}

The symbolic representation $\word(C^R)$ can be obtained from $\word(C)$ swapping min-channels with max-channels, and then selecting the lexicographically smallest representation for each type of word and for the whole sentence according to definitions \ref{defn:word} and \ref{defn:sentence}. Figure \ref{ex:reflection} shows two equivalent two-layer prefixes up to reflection, and the corresponding sentence representation.

Table \ref{tab:RHn} shows the cardinality of $|R(H_n)|$, $|R(T_n)|$ and $|R(T'_n)|$ for $n\leq 26$. For comparison, we also include the cardinality of $|R(G_n)|$, saturated prefixes $|R(S_n)|$ and saturated prefixes without reflections $|R(S'_n)|$ of interest in the optimal-depth sorting problem \cite{bundala2014optimal}.

\begin{theorem}
	For any $n\geq 3$, the set $R(T'_n)$ of two-layer comparator networks is a complete set of prefixes for the search of optimal sorting networks in size and depth.
\end{theorem}

\section{Results}
\label{sec:networks}

\begin{figure}[htb]
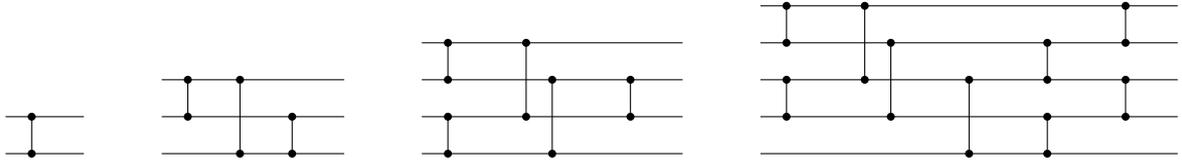

	\centering
	\begin{tabular}{cccc}
	\begin{sortingnetwork}{2}{0.7}
		\nodeconnection{ {0, 1}}
	\end{sortingnetwork}
	&
	\begin{sortingnetwork}{3}{0.7}
		\nodeconnection{ {0, 1}}
		\addtocounter{sncolumncounter}{1}
		\nodeconnection{ {0, 2}}
		\addtocounter{sncolumncounter}{1}
		\nodeconnection{ {1, 2}}
	\end{sortingnetwork}
	&
	\begin{sortingnetwork}{4}{0.7}
		\nodeconnection{ {0, 1}, {2, 3}}
		\addtocounter{sncolumncounter}{2}
		\nodeconnection{ {0, 2}}
		\nodeconnection{ {1, 3}}
		\addtocounter{sncolumncounter}{2}
		\nodeconnection{ {1, 2}}
	\end{sortingnetwork}
	&
	\begin{sortingnetwork}{5}{0.7}
		\nodeconnection{ {0, 1}, {2, 3}}
		\addtocounter{sncolumncounter}{2}
		\nodeconnection{ {0, 2}}
		\nodeconnection{ {1, 3}}
		\addtocounter{sncolumncounter}{2}
		\nodeconnection{ {2, 4}}
		\addtocounter{sncolumncounter}{2}
		\nodeconnection{ {1, 2}, {3, 4}}
		\addtocounter{sncolumncounter}{2}
		\nodeconnection{ {0, 1}, {2, 3}}
	\end{sortingnetwork}
	
	\end{tabular}
	\caption{Optimal sorting networks on $2$, $3$, $4$, and $5$ channels with $1$ layer and $1$ comparator, $3$ layers and $3$ comparators, $3$ layers and $5$ comparators, and $5$ layers and $9$ comparators, respectively.}
	\label{fig:optimal2}
\end{figure}

\begin{figure}[htb]
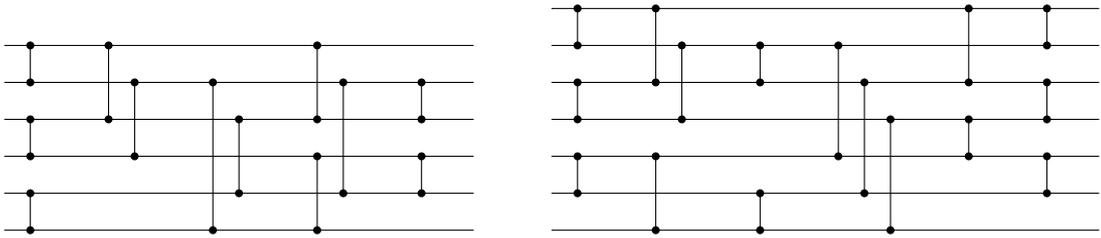

	\centering
	\begin{tabular}{cccc}
	\begin{sortingnetwork}{6}{0.7}
		\nodeconnection{ {0, 1}, {2, 3}, {4, 5}}
		\addtocounter{sncolumncounter}{2}
		\nodeconnection{ {0, 2}}
		\nodeconnection{ {1, 3}}
		\addtocounter{sncolumncounter}{2}
		\nodeconnection{ {1, 5}}
		\nodeconnection{ {2, 4}}
		\addtocounter{sncolumncounter}{2}
		\nodeconnection{ {0, 2}, {3, 5}}
		\nodeconnection{ {1, 4}}
		\addtocounter{sncolumncounter}{2}
		\nodeconnection{ {1, 2}, {3, 4}}
	\end{sortingnetwork}
	&
	\begin{sortingnetwork}{7}{0.7}
		\nodeconnection{ {0, 1}, {2, 3}, {4, 5}}
		\addtocounter{sncolumncounter}{2}
		\nodeconnection{ {0, 2}, {4, 6}}
		\nodeconnection{ {1, 3}}
		\addtocounter{sncolumncounter}{2}
		\nodeconnection{ {1, 2}, {5, 6}}
		\addtocounter{sncolumncounter}{2}
		\nodeconnection{ {1, 4}}
		\nodeconnection{ {2, 5}}
		\nodeconnection{ {3, 6}}
		\addtocounter{sncolumncounter}{2}
		\nodeconnection{ {0, 2}, {3, 4}}
		\addtocounter{sncolumncounter}{2}
		\nodeconnection{ {0, 1}, {2, 3}, {4, 5}}
	\end{sortingnetwork}
	\end{tabular}
	\caption{Optimal sorting networks on $6$ channels with $5$ layers and $12$ comparators, and on $7$ channels with $6$ layers and $16$ comparators.}
	\label{fig:optimal6}
\end{figure}

Using the propositional encodings of section \ref{sec:SAT} and standard SAT solvers\footnote{Unsatisfiability results were checked with $3$ SAT solvers: minisat, glucose, and cryptosat}, we can obtain optimal networks for $n \leq 10$ channels in a few seconds.
For $n \leq 9$ channels, there are networks that are optimal in both size and depth (Figures \ref{fig:optimal2}, \ref{fig:optimal6}, \ref{fig:optimal8}, and \ref{fig:optimal9}). 

For $n=10$ channels, optimal-depth sorting networks with 7 layers need a minimum of 31 comparators, while optimal-size sorting networks with 29 comparators require 8 layers (Figures \ref{fig:optimal10} and \ref{fig:optimal10b}).

\begin{theorem}
	The optimum size for a sorting network on $10$ channels of depth $7$ is $31$.
\end{theorem}

\begin{theorem}
	The optimum depth for a sorting network on $10$ channels with $29$ comparators is $8$.
\end{theorem}

For $n=11$ and $n=12$ channels, we use the results of section \ref{sec:symbolic} to fix the two first layers. The set of prefixes were also optimized with the evolutionary algorithm developed by Ehlers and Müller \cite{ehlers2015new} to reduce the number of variables in the resulting SAT formula.

The absolute minimal size $S(n)$ for $n=11$ channels is currently unknown. The lower bound on $S(11)$ is 33, but only networks with a minimum of $35$ comparators are known. Our depth-restricted results show that $35$ is the optimal size for sorting networks with $8$ or $9$ layers. For $n=12$ channels the lower bound of $S(n)$ is $37$ and the current upper bound $39$. In this case we obtain that optimal-depth sorting networks with $8$ layers need a minimum of $40$ comparators while for sorting networks with 9 layers the minimum is $39$ comparators. (Figures \ref{fig:optimal12} and \ref{fig:optimal12b})

For sorting networks on $11$ channels with $8$ layers, only 5 prefixes of the total of 403 prefixes in $R(T'_{11})$ give a network of size $35$ (Figures \ref{fig:optimal11}, \ref{fig:optimal11b}, \ref{fig:optimal11c}, \ref{fig:optimal11d}, \ref{fig:optimal11e}). Note that the first layer of the network on Figure \ref{fig:optimal11c} is not maximal, i.e., the prefix is not an element of the set $R(G_{11})$. For networks with $s\le34$ comparators and $d\le9$ layers all the propositional encodings with any of the 403 prefixes in $R(T'_{11})$ are unsatisfiable.

\begin{theorem}
	The optimum size for a sorting network on $11$ channels of depth $8$ or $9$ is $35$.
\end{theorem}

For depth-optimal sorting networks on $12$ channels with $8$ layers, the minimum number of comparators is $40$. This depth-restricted minimum size can be achieved with only 4 out of the 786 prefixes in $R(T'_{12})$. Figure \ref{fig:optimal12} is an example with only 5 comparators in the first layer. We need to increment the number of layers to $9$ to be able to obtain sorting networks on $12$ channels with $39$ comparators.

\begin{theorem}
	The optimum size for a sorting network on $12$ channels of depth $8$ is $40$.
\end{theorem}

For channels with more than $12$ channels of more than $9$ layers, the complexity of the SAT encoding begins to be out of the reach of current SAT solvers in the search of complete unsatisfiability results. However, we can use the proposed propositional encoding to obtain good networks for a given prefix in a few seconds or minutes, for example, a Green-type prefix \cite{knuth1998art}.

\section{Conclusions}

We have addressed the joint size and depth optimization of sorting networks, to obtain depth-restricted minimum size results for sorting networks on $n\le12$ channels. Our work extends the tools developed by Bundala et al. \cite{bundala2014optimal} for the search of depth-optimal networks. One of our contributions is the inclusion of size constraints in the propositional encoding of depth-restricted sorting networks. We have also addressed the symbolic representation of the general two-layer prefixes required in the proposed optimization problem.

We have shown that for $n=10$ channels, optimal-depth sorting networks with $7$ layers need a minimum of $31$ comparators, while optimal-size sorting networks with $29$ comparators require $8$ layers. The minimum size $S(n)$ for sorting networks on $n\ge11$ channels is currently unknown. However, our results show that $35$ is the minimum size for sorting networks on $11$ channels with $8$ or $9$ layers, and $40$ the minimum size for depth-optimal sorting networks on $12$ channels with $8$ layers. 

\section*{References}

\bibliographystyle{plain}
\bibliography{../sorting}

\begin{figure}[htb]
	\centering
	\begin{sortingnetwork}{8}{0.7}
		\nodeconnection{ {0, 1}, {2, 3}, {4, 5}, {6, 7}}
		\addtocounter{sncolumncounter}{2}
		\nodeconnection{ {0, 2}, {4, 6}}
		\nodeconnection{ {1, 3}, {5, 7}}
		\addtocounter{sncolumncounter}{2}
		\nodeconnection{ {0, 4}}
		\nodeconnection{ {1, 6}}
		\nodeconnection{ {2, 5}}
		\nodeconnection{ {3, 7}}
		\addtocounter{sncolumncounter}{2}
		\nodeconnection{ {1, 4}, {5, 6}}
		\addtocounter{sncolumncounter}{2}
		\nodeconnection{ {2, 4}}
		\nodeconnection{ {3, 5}}
		\addtocounter{sncolumncounter}{2}
		\nodeconnection{ {1, 2}, {3, 4}, {5, 6}}
	\end{sortingnetwork}
	\caption{Optimal sorting network on $8$ channels with $6$ layers and $19$ comparators.}
	\label{fig:optimal8}
\end{figure}

\begin{figure}[htb]
	\centering
	\begin{sortingnetwork}{9}{0.7}
		\nodeconnection{ {0, 1}, {2, 3}, {4, 5}, {6, 7}}
		\addtocounter{sncolumncounter}{2}
		\nodeconnection{ {0, 2}, {4, 6}}
		\nodeconnection{ {1, 3}, {5, 7}}
		\addtocounter{sncolumncounter}{2}
		\nodeconnection{ {0, 4}, {5, 6}}
		\nodeconnection{ {2, 8}}
		\nodeconnection{ {3, 7}}
		\addtocounter{sncolumncounter}{2}
		\nodeconnection{ {1, 6}}
		\nodeconnection{ {2, 5}}
		\nodeconnection{ {3, 8}}
		\addtocounter{sncolumncounter}{2}
		\nodeconnection{ {0, 2}, {3, 5}, {6, 8}}
		\nodeconnection{ {1, 4}}
		\addtocounter{sncolumncounter}{2}
		\nodeconnection{ {1, 2}, {3, 4}, {5, 6}, {7, 8}}
		\addtocounter{sncolumncounter}{2}
		\nodeconnection{ {2, 3}, {4, 5}}
	\end{sortingnetwork}
	\caption{Optimal sorting network on $9$ channels with $7$ layers and $25$ comparators.}
	\label{fig:optimal9}
\end{figure}

\begin{figure}[htb]
	\centering
	\begin{sortingnetwork}{10}{0.7}
		\nodeconnection{ {0, 1}, {2, 3}, {4, 5}, {6, 7}, {8, 9}}
		\addtocounter{sncolumncounter}{2}
		\nodeconnection{ {0, 2}, {4, 6}}
		\nodeconnection{ {1, 3}, {5, 7}}
		\addtocounter{sncolumncounter}{2}
		\nodeconnection{ {0, 4}, {5, 8}}
		\nodeconnection{ {1, 2}, {3, 7}}
		\nodeconnection{ {6, 9}}
		\addtocounter{sncolumncounter}{2}
		\nodeconnection{ {1, 6}}
		\nodeconnection{ {2, 8}}
		\nodeconnection{ {3, 9}}
		\nodeconnection{ {4, 5}}
		\addtocounter{sncolumncounter}{2}
		\nodeconnection{ {1, 4}, {8, 9}}
		\nodeconnection{ {2, 5}}
		\nodeconnection{ {3, 6}}
		\addtocounter{sncolumncounter}{2}
		\nodeconnection{ {0, 1}, {2, 4}, {6, 8}}
		\nodeconnection{ {3, 5}, {7, 9}}
		\addtocounter{sncolumncounter}{2}
		\nodeconnection{ {1, 2}, {3, 4}, {5, 6}, {7, 8}}
	\end{sortingnetwork}
	\caption{Optimal sorting network on $10$ channels with $7$ layers and $31$ comparators.}
	\label{fig:optimal10}
\end{figure}

\begin{figure}[htb]
	\centering
	\begin{sortingnetwork}{10}{0.7}
		\nodeconnection{ {0, 1}, {2, 3}, {4, 5}, {6, 7}, {8, 9}}
		\addtocounter{sncolumncounter}{2}
		\nodeconnection{ {0, 2}, {4, 6}}
		\nodeconnection{ {1, 3}, {5, 7}}
		\addtocounter{sncolumncounter}{2}
		\nodeconnection{ {0, 8}}
		\nodeconnection{ {2, 5}, {7, 9}}
		\addtocounter{sncolumncounter}{2}
		\nodeconnection{ {0, 4}, {6, 8}}
		\nodeconnection{ {1, 7}}
		\nodeconnection{ {3, 9}}
		\addtocounter{sncolumncounter}{2}
		\nodeconnection{ {1, 6}, {7, 8}}
		\nodeconnection{ {2, 4}}
		\nodeconnection{ {3, 5}}
		\addtocounter{sncolumncounter}{2}
		\nodeconnection{ {1, 2}, {3, 7}}
		\nodeconnection{ {4, 6}}
		\nodeconnection{ {5, 8}}
		\addtocounter{sncolumncounter}{2}
		\nodeconnection{ {2, 4}, {5, 7}}
		\nodeconnection{ {3, 6}}
		\addtocounter{sncolumncounter}{2}
		\nodeconnection{ {3, 4}, {5, 6}}
	\end{sortingnetwork}
	\caption{Optimal sorting network on $10$ channels with $8$ layers and $29$ comparators.}
	\label{fig:optimal10b}
\end{figure}

\begin{figure}[htb]
	\centering
	\begin{sortingnetwork}{11}{0.7}
		\nodeconnection{ {0, 1}, {2, 3}, {4, 5}, {6, 7}, {8, 9}}
		\addtocounter{sncolumncounter}{2}
		\nodeconnection{ {0, 6}, {8, 10}}
		\nodeconnection{ {1, 3}, {5, 7}}
		\nodeconnection{ {2, 4}}
		\addtocounter{sncolumncounter}{2}
		\nodeconnection{ {0, 8}, {9, 10}}
		\nodeconnection{ {1, 5}}
		\nodeconnection{ {3, 7}}
		\nodeconnection{ {4, 6}}
		\addtocounter{sncolumncounter}{2}
		\nodeconnection{ {0, 2}, {4, 9}}
		\nodeconnection{ {5, 8}}
		\nodeconnection{ {6, 10}}
		\addtocounter{sncolumncounter}{2}
		\nodeconnection{ {1, 4}, {6, 8}}
		\nodeconnection{ {2, 5}, {7, 10}}
		\nodeconnection{ {3, 9}}
		\addtocounter{sncolumncounter}{2}
		\nodeconnection{ {1, 2}, {3, 5}, {7, 9}}
		\nodeconnection{ {4, 6}}
		\addtocounter{sncolumncounter}{2}
		\nodeconnection{ {3, 4}, {5, 6}, {7, 8}}
		\addtocounter{sncolumncounter}{2}
		\nodeconnection{ {2, 3}, {4, 5}, {6, 7}, {8, 9}}
	\end{sortingnetwork}
	\caption{Optimal sorting network on $11$ channels with $8$ layers and $35$ comparators. Prefix $\sent{(012,12211221c)}$.}
	\label{fig:optimal11}
\end{figure}

\begin{figure}[htb]
	\centering
	\begin{sortingnetwork}{11}{0.7}
		\nodeconnection{ {0, 1}, {2, 3}, {4, 5}, {6, 7}, {8, 9}}
		\addtocounter{sncolumncounter}{2}
		\nodeconnection{ {0, 2}, {5, 6}, {8, 10}}
		\nodeconnection{ {1, 3}}
		\addtocounter{sncolumncounter}{2}
		\nodeconnection{ {0, 8}, {9, 10}}
		\nodeconnection{ {1, 7}}
		\nodeconnection{ {3, 6}}
		\nodeconnection{ {4, 5}}
		\addtocounter{sncolumncounter}{2}
		\nodeconnection{ {0, 4}, {7, 10}}
		\nodeconnection{ {1, 5}}
		\nodeconnection{ {2, 9}}
		\nodeconnection{ {3, 8}}
		\addtocounter{sncolumncounter}{2}
		\nodeconnection{ {1, 2}, {3, 4}, {5, 9}}
		\nodeconnection{ {6, 10}}
		\nodeconnection{ {7, 8}}
		\addtocounter{sncolumncounter}{2}
		\nodeconnection{ {1, 3}, {4, 5}, {6, 8}}
		\nodeconnection{ {2, 7}}
		\addtocounter{sncolumncounter}{2}
		\nodeconnection{ {2, 4}, {5, 7}}
		\nodeconnection{ {6, 9}}
		\addtocounter{sncolumncounter}{2}
		\nodeconnection{ {2, 3}, {4, 5}, {6, 7}, {8, 9}}
	\end{sortingnetwork}
	\caption{Optimal sorting network on $11$ channels with $8$ layers and $35$ comparators. Prefix $\sent{(012,1212,1221c)}$.}
	\label{fig:optimal11b}
\end{figure}

\begin{figure}[htb]
	\centering
	\begin{sortingnetwork}{11}{0.7}
		\nodeconnection{ {0, 1}, {2, 3}, {4, 5}, {6, 7}, {8, 9}}
		\addtocounter{sncolumncounter}{2}
		\nodeconnection{ {0, 2}, {4, 6}, {8, 10}}
		\nodeconnection{ {1, 3}, {5, 7}}
		\addtocounter{sncolumncounter}{2}
		\nodeconnection{ {0, 4}, {9, 10}}
		\nodeconnection{ {2, 5}}
		\nodeconnection{ {3, 7}}
		\addtocounter{sncolumncounter}{2}
		\nodeconnection{ {1, 5}, {6, 9}}
		\nodeconnection{ {2, 8}}
		\nodeconnection{ {4, 10}}
		\addtocounter{sncolumncounter}{2}
		\nodeconnection{ {0, 2}, {3, 10}}
		\nodeconnection{ {1, 6}}
		\nodeconnection{ {4, 8}}
		\nodeconnection{ {5, 9}}
		\addtocounter{sncolumncounter}{2}
		\nodeconnection{ {2, 4}, {5, 8}}
		\nodeconnection{ {3, 6}, {7, 10}}
		\addtocounter{sncolumncounter}{2}
		\nodeconnection{ {1, 4}, {6, 8}}
		\nodeconnection{ {3, 5}, {7, 9}}
		\addtocounter{sncolumncounter}{2}
		\nodeconnection{ {1, 2}, {3, 4}, {5, 6}, {7, 8}}
	\end{sortingnetwork}
	\caption{Optimal sorting network on $11$ channels with $8$ layers and $35$ comparators. Prefix $\sent{(012,1221c,1221c)}$.}
	\label{fig:optimal11c}
\end{figure}

\begin{figure}[htb]
	\centering
	\begin{sortingnetwork}{11}{0.7}
		\nodeconnection{ {0, 1}, {2, 3}, {4, 5}, {8, 9}}
		\addtocounter{sncolumncounter}{2}
		\nodeconnection{ {0, 2}, {4, 7}, {8, 10}}
		\nodeconnection{ {1, 3}, {5, 6}}
		\addtocounter{sncolumncounter}{2}
		\nodeconnection{ {0, 8}, {9, 10}}
		\nodeconnection{ {1, 7}}
		\nodeconnection{ {3, 6}}
		\nodeconnection{ {4, 5}}
		\addtocounter{sncolumncounter}{2}
		\nodeconnection{ {0, 4}, {7, 10}}
		\nodeconnection{ {1, 5}}
		\nodeconnection{ {2, 9}}
		\nodeconnection{ {3, 8}}
		\addtocounter{sncolumncounter}{2}
		\nodeconnection{ {1, 2}, {3, 4}, {5, 9}}
		\nodeconnection{ {6, 10}}
		\nodeconnection{ {7, 8}}
		\addtocounter{sncolumncounter}{2}
		\nodeconnection{ {1, 3}, {4, 7}}
		\nodeconnection{ {2, 5}, {6, 9}}
		\addtocounter{sncolumncounter}{2}
		\nodeconnection{ {2, 4}, {5, 7}}
		\nodeconnection{ {6, 8}}
		\addtocounter{sncolumncounter}{2}
		\nodeconnection{ {2, 3}, {4, 5}, {6, 7}, {8, 9}}
	\end{sortingnetwork}
	\caption{Optimal sorting network on $11$ channels with $8$ layers and $35$ comparators.  Prefix $\sent{(012,0120,1221c)}$.}
	\label{fig:optimal11d}
\end{figure}

\begin{figure}[htb]
	\centering
	\begin{sortingnetwork}{11}{0.7}
		\nodeconnection{ {0, 1}, {2, 3}, {4, 5}, {6, 7}, {8, 9}}
		\addtocounter{sncolumncounter}{2}
		\nodeconnection{ {0, 2}, {4, 10}}
		\nodeconnection{ {1, 3}, {5, 7}}
		\nodeconnection{ {6, 8}}
		\addtocounter{sncolumncounter}{2}
		\nodeconnection{ {4, 6}, {8, 10}}
		\nodeconnection{ {5, 9}}
		\addtocounter{sncolumncounter}{2}
		\nodeconnection{ {0, 5}, {7, 9}}
		\nodeconnection{ {1, 8}}
		\nodeconnection{ {2, 6}}
		\nodeconnection{ {3, 10}}
		\addtocounter{sncolumncounter}{2}
		\nodeconnection{ {0, 4}, {6, 9}}
		\nodeconnection{ {1, 2}, {3, 5}, {7, 8}}
		\addtocounter{sncolumncounter}{2}
		\nodeconnection{ {1, 4}, {5, 8}, {9, 10}}
		\nodeconnection{ {2, 3}, {6, 7}}
		\addtocounter{sncolumncounter}{2}
		\nodeconnection{ {2, 4}, {5, 7}, {8, 9}}
		\nodeconnection{ {3, 6}}
		\addtocounter{sncolumncounter}{2}
		\nodeconnection{ {3, 4}, {5, 6}, {7, 8}}
	\end{sortingnetwork}
	\caption{Optimal sorting network on $11$ channels with $8$ layers and $35$ comparators. Prefix $\sent{(0122112,1221c)}$.}
	\label{fig:optimal11e}
\end{figure}

\begin{figure}[htb]
	\centering
	\begin{sortingnetwork}{12}{0.7}
		\nodeconnection{ {0, 1}, {2, 3}, {4, 5}, {6, 7}, {8, 9}}
		\addtocounter{sncolumncounter}{2}
		\nodeconnection{ {0, 2}, {4, 6}, {8, 11}}
		\nodeconnection{ {1, 3}, {5, 7}, {9, 10}}
		\addtocounter{sncolumncounter}{2}
		\nodeconnection{ {0, 8}}
		\nodeconnection{ {1, 2}, {3, 11}}
		\nodeconnection{ {4, 9}}
		\nodeconnection{ {5, 6}, {7, 10}}
		\addtocounter{sncolumncounter}{2}
		\nodeconnection{ {0, 4}, {5, 8}, {10, 11}}
		\nodeconnection{ {1, 6}}
		\nodeconnection{ {2, 7}}
		\nodeconnection{ {3, 9}}
		\addtocounter{sncolumncounter}{2}
		\nodeconnection{ {1, 3}, {4, 5}, {6, 9}}
		\nodeconnection{ {2, 8}}
		\nodeconnection{ {7, 10}}
		\addtocounter{sncolumncounter}{2}
		\nodeconnection{ {1, 4}, {7, 8}, {9, 10}}
		\nodeconnection{ {2, 5}}
		\nodeconnection{ {3, 6}}
		\addtocounter{sncolumncounter}{2}
		\nodeconnection{ {2, 4}, {6, 7}, {8, 9}}
		\nodeconnection{ {3, 5}}
		\addtocounter{sncolumncounter}{2}
		\nodeconnection{ {3, 4}, {5, 6}, {7, 8}}
	\end{sortingnetwork}
	\caption{Optimal sorting network on $12$ channels with $8$ layers and $40$ comparators.}
	\label{fig:optimal12}
\end{figure}

\begin{figure}[htb]
	\centering
	\begin{sortingnetwork}{12}{0.7}
		\nodeconnection{ {0, 1}, {2, 3}, {4, 5}, {6, 7}, {8, 9}, {10, 11}}
		\addtocounter{sncolumncounter}{2}
		\nodeconnection{ {0, 10}}
		\nodeconnection{ {1, 3}, {5, 7}, {9, 11}}
		\nodeconnection{ {2, 4}, {6, 8}}
		\addtocounter{sncolumncounter}{2}
		\nodeconnection{ {0, 2}, {3, 11}}
		\nodeconnection{ {1, 9}}
		\nodeconnection{ {4, 10}}
		\addtocounter{sncolumncounter}{2}
		\nodeconnection{ {0, 6}, {7, 11}}
		\nodeconnection{ {1, 5}, {8, 10}}
		\nodeconnection{ {2, 4}}
		\nodeconnection{ {3, 9}}
		\addtocounter{sncolumncounter}{2}
		\nodeconnection{ {1, 6}, {7, 10}}
		\nodeconnection{ {2, 8}}
		\nodeconnection{ {5, 9}}
		\addtocounter{sncolumncounter}{2}
		\nodeconnection{ {1, 2}, {3, 7}, {9, 10}}
		\nodeconnection{ {4, 6}}
		\nodeconnection{ {5, 8}}
		\addtocounter{sncolumncounter}{2}
		\nodeconnection{ {2, 4}, {7, 9}}
		\nodeconnection{ {3, 6}}
		\addtocounter{sncolumncounter}{2}
		\nodeconnection{ {3, 5}, {6, 8}}
		\addtocounter{sncolumncounter}{2}
		\nodeconnection{ {3, 4}, {5, 6}, {7, 8}}
	\end{sortingnetwork}
	\caption{Optimal sorting network on $12$ channels with $9$ layers and $39$ comparators.}
	\label{fig:optimal12b}
\end{figure}

\end{document}